# Ranking RDF Instances in Degree-decoupled RDF Graphs

Elisa S. Menendez[1,2], Marco A. Casanova[1], Mohand Boughanem[2],
Luiz André P. Paes Leme[3]

[1]Department of Informatics – Pontifical Catholic University of Rio de Janeiro, RJ, Brazil
[2]Institute de Recherche en Informatique de Toulouse, Toulouse, France
[3]Fluminense Federal University, Niterói, RJ, Brazil
{emenendez,casanova}@inf.puc-rio.br, boughanem@irit.fr,
lleme@id.uff.br

**Abstract.** In the last decade, RDF emerged as a new kind of standardized data model, and a sizable body of knowledge from fields such as Information Retrieval was adapted to RDF graphs. One common task in graph databases is to define an importance score for nodes based on centrality measures, such as PageRank and HITS. The majority of the strategies highly depend on the degree of the node. However, in some RDF graphs, called degree-decoupled RDF graphs, the notion of importance is not directly related to the node degree. Therefore, this work first proposes three novel node importance measures, named *InfoRank I, II* and *III*, for degree-decoupled RDF graphs. It then compares the proposed measures with traditional PageRank and other familiar centrality measures, using with an IMDb dataset.

## 1  Introduction

In the last decade, driven by the Semantic Web and Linked Data principles, RDF emerged as a new kind of data model that organizes data as triples, whose collection induces an RDF graph. This kind of modeling provides more flexibility to describe resources and follows W3C standardized formats and ontologies. A sizable body of knowledge from fields such as Information Retrieval, Semantic Relatedness, Named Entity Recognition, etc. was then adapted to RDF graphs.

One common task among these fields is the ranking of entities based on their importance. This is typically computed using centrality measures, in which the notion of importance is based on how connected a given entity, represented by a node, is to the rest of the graph. PageRank [5] and HITS [13] are the most popular centrality measures used in Information Retrieval to compute the importance of Web pages. Therefore, considerable research has been devoted to adapt one of these measures to RDF graphs.

The majority of the related work test their strategies using some RDF graph that reflects Web pages and their links, extracted from DBpedia[1] [7, 9-11, 14, 16], for example, or using some dataset about co-authorships of research papers, with data from DBLP[2] [2, 7, 20], for example. We argue that PageRank or HITS variations work well for these types of datasets because the incoming or outgoing edges actually indicate the relevance of a resource. In the Web, it is reasonable that a Web page (or node) with several incoming edges is more important than a Web page with a few incoming edges. Likewise, in a research publications dataset, made of just the citation relationship, a paper with many citations (incoming edges) is usually more important than a paper with a few citations.

However, some datasets may follow a schema in which the incoming or outgoing edges of a node do not directly indicate its importance with respect to any existing node relationship or, at least, make it be hard to detect which relationship would express this notion. With such datasets, the whole graph must be processed, and traditional measures usually fail to compute the importance of a node. As an example, in the IMDb dataset, a TV Series that has been on the air for a long time has many episodes linked to it (e.g. "General Hospital" has 14,000 episodes) and there is no other relationship that would connect TV Series and that would induce a notion of importance. Hence, a traditional PageRank algorithm would assign a higher score to these TV Series than to those that have just a few episodes (e.g. "Friends" with 236 episodes). Of course, we could manually assign weights to the links in order to capture their semantics, and use a Weighted PageRank or HITS Algorithm, as in [2, 6, 17]. However, many argue that the manual assignment of weights to links is bothersome and subjective. Thus, other works focus on learning strategies based on user's feedbacks [1, 14, 16].

In addition to the difficulty of detecting relationships that express the importance of a graph node, it would be interesting to define new importance measures that do not require a preliminary filtering step that eliminates unwanted relationships that would distort traditional centrality measures.

The first contribution of this work is the definition of three node importance measures, named *InfoRank I, InfoRank II and InfoRank III*, for *degree-decoupled* RDF graphs, that is, RDF graphs in which the importance of a node is not directly related to its degree. The proposed node importance measures are combinations of three intuitions: (I) important things have lots of information about them; (II) important things are surrounded by other important things; (III) few good friends are better than many acquaintances. They require neither the manual assignment of link weights nor a training set to use as input to a learning algorithm. Furthermore, InfoRank I is straightforward to compute and InfoRank II and InfoRank III have a complexity comparable to PageRank.

The second contribution of this work is an evaluation of the proposed node importance measures, based on an IMDb dataset, vis-à-vis variations of different centrality measures – PageRank, Weighted PageRank, Eigenvector Centrality. The evalua-tion compares the ranks induced by these measures with a gold standard for movies, series and actors based on user ratings extracted from the IMDb Web site. Furthermore, since our goal was to capture a notion of popular works rather than notorious ones, we performed another evaluation using the sum of the number of user votes, also extracted from the IMDb Web site. Our goal was to test if we could achieve a good popularity measure (similar to user ratings and votes) using only the structure of the graph. We show that *InfoRank I* achieves the best result considering the gold standards and is very efficient, while *InfoRank III*, that is, using the three intuitions, achieves the best result considering user's votes.

---

[1] http://dbpedia.org/sparql
[2] http://dblp.uni-trier.de

The remainder of the paper is organized as following. Section 2 reviews related work. Section 3 summarizes some background concepts about RDF and centrality measures. Section 4 presents the InfoRank measures. Section 5 presents the evaluation using an IMDb dataset. Finally, Section 6 contains the conclusions.

## 2    Related Work

ObjectRank [2] was one of the first proposals to adapt PageRank to graph databases. The authors transformed the structure of a relational database (RDB) to a graph, using foreign keys as links between entities, and them applied PageRank with manual link weights to compute a global importance score for entities. The authors evaluated their strategy using the DBLP dataset.

TripleRank [7] adapts the HITS algorithm to rank RDF triples and performs the evaluation using both DBpedia and DBLP. The main contribution is a novel representational model for RDF graphs based on a 3D tensor. As in HITS, they use the concepts of *authorities* and *hubs*, which are basically nodes with high in-degree and out-degree, respectively.

More recently, FORK [14] adapted ObjectRank to Linked Data. The main contribution of the work is a learning algorithm for link weights based on user's relevance feedbacks, instead of the manual assignment of weights. The authors evaluated their strategy using DBpedia and results showed that FORK achieves the best ranking method when compared to baseline approaches.

As mentioned on the introduction, DBpedia and DBLP are highly influenced by link semantics: DBLP through citations links, and DBpedia through links derived from Wikipedia, such as, *wikiPageRedirects, wikiPageDisambiguates, primaryTopic, etc*. Furthermore, in the LOD[3] cloud, DBpedia has many incoming links from other RDF datasets.

For further references that focus on ranking strategies for *degree-dependent* datasets, such as DBpedia or DBLP, we refer the reader to the surveys in [3, 18, 22]. We continue our discussion with some alternative strategies that do not highly depend on node degree.

The work in [8] proposes the use of closeness centrality on undirected graphs and evaluates the strategy using three datasets, CIA Factbook, Terrorist Ontology and Wine Ontology. The authors compare their strategy with a ranking using the number of incoming edges. The problem with closeness centrality is that it is not efficient for large RDF graphs.

---

[3] http://lod-cloud.net

Although the work presented in [12] is not specific to RDF graphs, it proposes the *degree decoupled PageRank* technique that penalizes or boosts the importance of the node degree in recommendation graphs, depending on the domain characteristics. They argue that, in some contexts, the importance of the node can be inversely proportional to its degree. The authors performed an evaluation using graphs extracted from IMDb, Last.fm, DBLP and Epinions. From results for the IMDb dataset, they noticed that, for a movie recommendation graph, traditional PageRank performs better; however, for an actor recommendation graph, the node degree actually needs to be penalized. They argue that an actor that plays in a large number of movies may be a non-discriminating ("B movie") actor, whereas an actor with relatively few movies may be a more discriminating ("A movie") actor.

## 3    Background

### 3.1    RDF Basics

We first summarize some basic concepts pertaining to the *Resource Description Framework* (RDF)[4].

An *Internationalized Resource Identifier* (*IRI*) is a global identifier that denotes a resource. We will use the term IRI and resource interchangeably. A *literal* is a basic value, such as strings, numbers, dates, etc. An *RDF term* is either an IRI or a literal. We will use $R$ to denote the set of all IRIs and $L$ to denote the set of all literals.

RDF models data as triples of the form *(s,p,o)*, where *s* and *p* are IRIs and *o* may be an IRI or a literal; we say that *s* is the *subject*, *p* is the *predicate* and *o* is the *object* of the triple. An RDF triple *(s,p,o)* says that some relationship, indicated by *p*, holds from the subject *s* to the object *o*. When the object of triple is an IRI we say that *p* is an *object property* and when the object is a literal we say that *p* is a datatype property.

A set $T$ of RDF triples is equivalent to a labeled graph $G$ such that the set of nodes of $G$ is the set of RDF terms that occur as subject or object of the triples in $T$ and there is a directed edge *(s,o)* in $G$ labeled with *p* iff the triple *(s,p,o)* occurs in $T$. Therefore, we will use the concepts of set of RDF triples and RDF graph interchangeably.

### 3.2    Centrality Measures

Centrality measures have as goal to identify the most important or central node in a graph, depending on what importance means. A simple way to compute the centrality of a node is just to analyze its degree. However, this returns a local measure of centrality, whereas in some contexts a global analysis of the graph is preferable. For instance, the *Betweenness Centrality* counts the number of shortest paths going through a node; hence it is able to identify important connectors in a graph. The *Closeness Centrality* measures the average distance from a node to all other nodes, hence the more central a node is, the closer it is to all other nodes.

---

[4] http://www.w3.org/TR/rdf11-concepts/

Other types of centrality measures try to capture the idea that "it is not about what you know, but who you know". That is, the notion of importance is given by how well connected a node is to other important nodes.

PageRank [5] is the most popular centrality measure of this type. Using the hyperlink structure of the Web, the basic idea is that Web pages that are linked from many other Web pages are probably high-quality pages. If a Web page $P$ has links from other high-quality Web pages, for instance Wikipedia, then that is a further indication that it is likely to be worth looking at $P$.

PageRank can be computed using an iterative strategy, named the *Power Iteration* method. Let $G = (V, E)$ be a directed graph and $PR(v, i)$ be the PageRank score calculated at iteration $i$. First, we initialize all scores with the same value; then, for $0 < i < x$, we iterate until the computation of the centrality score converges or exceeds the maximum number of possible iterations $x$.

**Measure 1.** PageRank

$$PR(v, 0) = \frac{1}{n} \tag{1}$$

$$PR(v, i) = \frac{1-\alpha}{n} + \alpha \sum_{t \in M_I(v)} \frac{PR(t, i-1)}{d_O(t)} \tag{2}$$

where $\alpha$ is a dumping factor (usually set to 0.85), $n$ is the total number of nodes in $G$, $M_I(v)$ is the set of nodes that have a link to $v$ and $d_O(t)$ is the number of outgoing links from $t$.

One variant of PageRank is the use of link weights to give more importance for certain types of links. The Weighted PageRank $PW(v,i)$, for iteration $i$, is defined as follows.

**Measure 2.** Weighted PageRank

$$PW(v, 0) = \frac{1}{n} \tag{3}$$

$$PW(v, i) = \frac{1-\alpha}{n} + \alpha \sum_{t \in M_I(v)} \frac{PW(t, i-1)}{d_O(t)} * lw(v, t) \tag{4}$$

where $lw(v,t)$ is the weight of a link $(v, t) \in E$.

PageRank is actually inspired by an older measure, called *Eigenvector Centrality*, which can also be computed using the Power Iteration method. First, we initialize the weights as in PageRank, and then we iterate until it converges or exceeds the maximum number of possible iterations.

6**Measure 3.** Eigenvector Centrality

$$EC(v, 0) = \frac{1}{n} \qquad (5)$$

$$EC(v, i+1) = EC(v, i) + \sum_{t \in M(v)} EC(t, i) \qquad (6)$$

where $M(v)$ is the set of neighbors of $v$. Then, the values of $EC(v, i+1)$ are normalized as following:

$$norm = \sqrt{\sum_{v \in G} EC(v, i+1)^2} \qquad (7)$$

$$EC(v, i+1) = \frac{EC(v, i+1)}{norm} \qquad (8)$$

### 3.3 Precision and Average Precision

When applied to a graph, a centrality or importance measure induces a ranked list $L$ of the set of nodes of the graph, which can be compared to a *golden standard* list $S$. In this section, we therefore recall the definitions of precision and average precision [19] for the ranked lists, which we will use to compare the measures.

Let $S$ be a list of documents, considered as the *golden standard*, and let $R$ be the set of all documents in $S$. A document $d$ is *relevant* iff $d \in R$. Let $L$ be a list of documents.

The *precision at position k* of $L$ with respect to $S$ is defined as:

$$P(k) = \frac{|R \cap retrieved\ (k)|}{|retrieved\ (k)|} \qquad (9)$$

where *retrieved*($k$) is the set of all documents in $L$ until position $k$.

The *average precision* of $L$, denoted $AP_L$, with respect to $S$ is defined as:

$$AP_L = \frac{1}{|R|} \sum_{k=1}^{n} relevance(k) * P(k) \qquad (10)$$

where *relevance*($k$) is an indicator function that returns *1*, if the document at position $k$ is relevant, and *0*, otherwise. Notice that the average precision of the golden standard $S$ is $AP_S = 1$, which is the target performance of a centrality measure.

## 4 The InfoRank Measures

### 4.1 The intuition behind the InfoRank Measures

Following the intuition that "important things have lots of information about them" and observing the way that RDF graphs are modeled in our test dataset (IMDb), we notice that important nodes usually have more datatype properties (information) than less important nodes. When comparing with a relational database, it would be the same as saying that important tuples are those with known attributes (i.e. different from *null*), which are translated to datatype properties in an RDF graph. Furthermore, when modeling RDF graphs, the normal forms are usually not applied; hence, all information about an entity can be directly connected to the node.

The second intuition that we follow is inspired by PageRank and says that "important things are surrounded by other important things". For instance, the movie "Titanic" has links through object properties with actors "Kate Winslet" and "Leonardo Dicaprio", which are also important nodes in the graph. As in [8], we agree that in RDF datasets the direction of an object property does not have the same meaning as a Web hyperlink since a property is often found in its inverse form (e.g. directedBy/hasDirector). Given that, we treat an RDF graph as undirected and consider all neighbors of a node (i.e. all other nodes that have an object property linked to it) when propagating the importance.

However, our goal is to further improve this intuition, following a third one that says "few good friends are better than many acquaintances". As discussed in the introduction, the typical centrality measures are highly dependent on the degree of the node. In our work, we do not want to boost (or penalize) the degree importance, but we focus on a strategy that favors the quality of neighbors, rather than their quantity, that is, we prefer an approach that captures the notion that "10 popular neighbors (or friends) are better than 1,000 unpopular neighbors".

### 4.2 Computation of the InfoRank Measures

Given the three intuitions, we propose three versions of InfoRank, defined in the following.

Let $T$ be the set of triples consisting only of ABox data, that is, $T$ contains no schema information. Let $R$ be the set of all *instances*, that is, $v \in R$ iff $(v, rdf{:}type, c) \in T$, and $L$ the set of all literals. We say that $t$ is a *neighbor* of $v$ iff $t \in R$ and $(v, p, t) \in T$ or $(t, p, v) \in T$. Let $G = (V, E)$ be the graph induced by the relationship between instances and their neighbors (note that $V \subseteq R$). Furthermore, let $dtp(v)$ denote the number of datatype properties of a node $v \in R$.

The first version of InfoRank, named *InfoRank I*, tries to capture our first intuition and is simply defined as follows:

**Measure 4.**  InfoRank I

$$w(v) = dtp(v) \bigg/ \sum_{v \in R} dtp(t) \tag{11}$$

In order to capture our first and second intuitions together, we define *InfoRank II*, which extends the Eigenvector Centrality by using *InfoRank I* as node weights.

**Measure 5.** InfoRank II

$$IR2(v, 0) = w(v) \tag{12}$$

$$IR2(v, i+1) = IR2(v, i) + \sum_{t \in M(v)} IR2(t, i) * \big(w(v) + w(t)\big) \tag{13}$$

$$norm = \sqrt{\sum_{v \in R} IR2(v, i+1)^2} \tag{14}$$

$$IR2(v, i+1) = \frac{IR2(v, i+1)}{norm} \tag{15}$$

Finally, to capture the three intuitions, we define *InfoRank III*, which extends *InfoRank II* by limiting the set of neighbors $M(v)$ to only the top ones. Let $M_z(v, i)$ be the top $z$ neighbors of a node $v$ computed in the previous iteration. For example, for z=100, we use the top 100 neighbors of $v$ ordered by their InfoRank value in iteration $i$.

**Measure 6.** InfoRank III

$$IR3(v, 0) = w(v) \tag{16}$$

$$IR3(v, i+1) = IR3(v, i) + \sum_{t \in M_z(v)} IR3(t, i) * \big(w(v) + w(t)\big) \tag{17}$$

$$norm = \sqrt{\sum_{v \in R} IR3(v, i+1)^2} \tag{18}$$

$$IR3(v, i+1) = \frac{IR3(v, i+1)}{norm} \tag{19}$$

An example of how to compute InfoRank is as follows.

**Example.** Consider the simple graph shown in Figure 1 with IRIs denoted in oval and literals denoted in dashed boxes. Note that node *D* has a degree of 34, since it has as neighbors nodes $D_1$ to $D_{33}$ and *A*. Hence, *D* would receive the highest score using traditional PageRank or Eigenvector Centrality due to the high degree, when compared with other nodes. But, considering that this is not a *degree-dependent* dataset and following our intuitions, we prefer that *A*, *B* and *C* have higher scores than *D*.

Note that *InfoRank I* (i.e. counting literals) gives the highest scores to nodes *B* and *C*. However, we would like *A* to also have a high score, since it is very central and it is surrounded by important neighbors. Using *InfoRank II* is still not enough to achieve our goal, since *D* has a high degree. Finally, *InfoRank III* achieves our goal for this example. Table 1 shows the count of datatype properties (*dtp*) and the result of some iterations for *InfoRank III* according to the Power Method and using $z = 10$. Note that, to help the explanation, we round the numbers after each iteration.

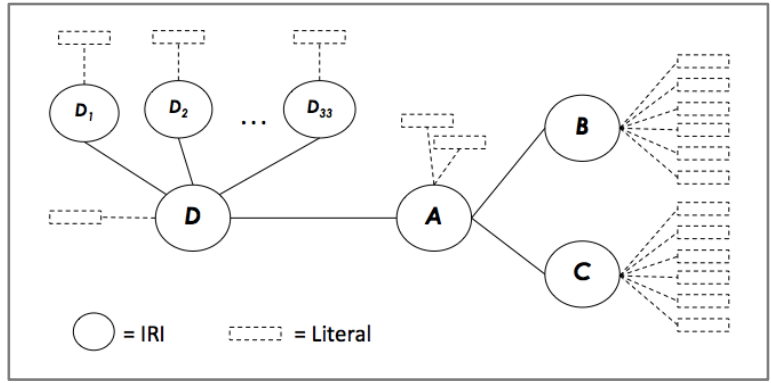

**Fig 1**. Graph example.

**Table 1**. InfoRank III iterations example.

| Node | dtp | i = 0 | i = 1 | i = 2 | i = 3 | i = 17 |
|---|---|---|---|---|---|---|
| A | 2 | 0,042 | 0,351 | 0,465 | **0,54** | **0,679** |
| B, C | 6 | **0,125** | **0,544** | **0,518** | 0,497 | 0,459 |
| D | 1 | 0,021 | 0,13 | 0,16 | 0,184 | 0,24 |
| $D_1 \ldots D_{33}$ | 1 | 0,021 | 0,09 | 0,082 | 0,074 | 0,043 |

In iteration $i = 2$, score *IR3* (before the normalization) of node *D* is given by:

$$IR3(D, 2) = 0,13 + (0,351 * 0,063) + (0,09 * 0,042 * 9)$$

Note that the last parenthesis represents the sum of the neighbors $D_1 \ldots D_9$, since we only use the top 10 neighbors of node *D*, and *A* is the first one on the list according to the *IR3* scores of last iteration ($i = 1$).

Finally, we point out that, after iteration 3, *A* achieves the first place, followed by *B* and *C*, and *D* is only in the fourth position. The example converged after 17 iterations.

## 5   Evaluation

In order to evaluate our strategy, we downloaded the relational IMDb[5] dataset in MySQL and used Oracle 12c to transform it to RDF via R2RML. An overview of the RDF schema is shown in Figure 2. The resulting schema has 22 classes, 92 distinct datatype properties and 32 distinct object properties. The resulting RDF dataset has a total of 180,704,900 triples, of which 30,154,661 are relationships between instances and classes (through rdf:type property), 80,326,633 are relationships between instances and instances (through object properties), and 70,223,606 are relationships between instances and literals (through datatype properties). Recall that, when computing the centrality measures, we do not consider the instance/class relationships and we use the literals as weights, instead of considering them as nodes. Hence, we end up with a graph with 29,036,914 nodes (i.e. distinct instances). Furthermore, note that an RDF graph can be multigraph, that is, it can have more than one edge between two nodes (e.g. a person can be the director and the writer of a movie). However, when calculating the centrality measures, we consider only distinct relationships, hence, we end up with 78,511,197 edges.

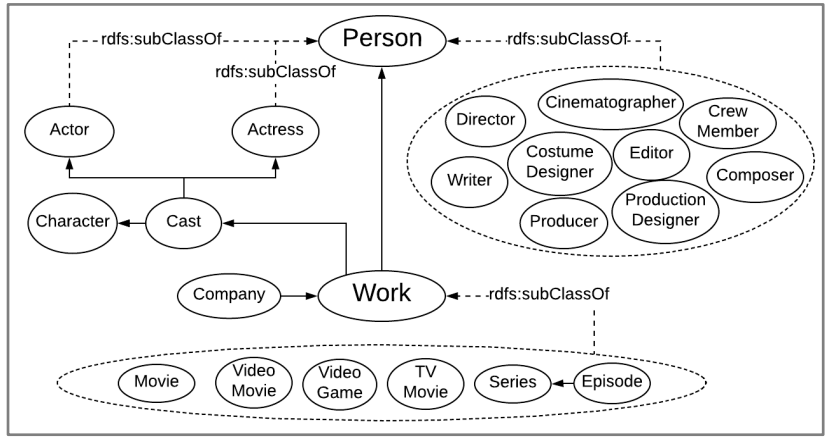

**Fig 2**. IMDb Schema.

As the gold standard, we used the IMDb lists of 100 top rated movies[6] and series[7] and 100 greatest actors[8], denoted $S_{Movies}$, $S_{Series}$ and $S_{Actors}$. Each centrality or node importance measure induces a list of movies, and, therefore, has an average precision $AP_{Movies}$ with respect to the golden standard for movies. Likewise, we have an average precision for TV series, denoted $AP_{Series}$, and for actors, denoted $AP_{Actors}$. The *combined mean average precision*, denoted *CMAP*, with respect to the gold standards for movies, TV series and actors is defined as:

$$CMAP = \frac{AP_{Movies} + AP_{Series} + AP_{Actors}}{3} \qquad (20)$$

---

[5] https://sites.google.com/site/ontopiswc13/home/imdb-mo
[6] https://www.imdb.com/chart/top
[7] https://www.imdb.com/list/ls008957859/
[8] https://www.imdb.com/list/ls050274118/

Table 2 shows the average precision for the centrality and node importance measures we consider for movies, TV series and actors, as well as their combined MAP. Table 2 also shows the time in minutes and the number of iterations required to converge. We implemented a version of the Power Iteration method using SPARQL and SQL queries, since our triples were stored in an Oracle database. For these experiments, we used Oracle 12c, running on a 2x deca-core Intel(R) Xeon(R) CPU E5-2640 v4 @ 2.40GHz, 128GB RAM, 32KB Cache L1.

**Table 2**. Results for Gold Standard.

| Measures | $AP_{Movies}$ | $AP_{Series}$ | $AP_{Actors}$ | CMAP | Time (min) | Num Iter |
|---|---|---|---|---|---|---|
| In-degree | 0,553 | 0,011 | 0,3333 | 0,299 | - | - |
| Out-degree | 0,530 | 0,158 | 0 | 0,229 | - | - |
| Degree | 0,540 | 0,026 | 0,3333 | 0,300 | - | - |
| PageRank | 0,264 | 0 | 0,2500 | 0,171 | 59 | 48 |
| Weighted PageRank | 0,350 | 0 | 0,3333 | 0,228 | 22 | 19 |
| Eigenvector | * | * | * | * | * | * |
| InfoRank I | 0,424 | 0,350 | **0,6182** | **0,464** | - | - |
| InfoRank II | 0,587 | 0,389 | 0,1470 | 0,374 | 97 | 63 |
| InfoRank III - 10 | 0,493 | 0,357 | 0,1525 | 0,334 | 63 | 24 |
| InfoRank III - 100 | 0,529 | **0,512** | 0,1255 | 0,389 | 81 | 33 |
| InfoRank III - 1000 | **0,605** | 0,449 | 0,1301 | 0,395 | 89 | 35 |

\* No convergence in 200 iterations

Consider first the measures that take into the account the node degrees. The *In-degree* measure orders the instances by the number of incoming edges in descending order. The *Out-degree* and *Degree* measures are likewise defined, except that they count only the object properties.

These measures have good average precision for movies due to the properties *references/remake_of*, which is a good indicator of importance, since great movies are frequently referenced by other works. Indeed, the gold standard list has a large number of classic movies, such as The Godfather (1972), Casablanca (1942), Psychosis (1960), etc., which have more references throughout time than more recent ones. However, they have poor average precision for TV series, as a consequence of the problem mentioned in the introduction: TV series that have been on the air for a long time have a large number of episodes linked to it (e.g. "General Hospital" has 14,000 episodes). For actors, the *Out-degree* has a negligible performance since the instances have no outgoing edges. Hence, they have just a reasonable combined MAP.

PageRank has a poor average precision for movies and actors, and negligible average precision for TV series. Hence, it has a very poor combined MAP.

For Weighted PageRank, we tried to capture the semantics of the properties by assigning weights from 0 to 1 according to the type of property and using a directed graph. For example, properties *references*, *remake_of*, *version_of*, etc., were assigned weight 1.0, and properties *episode_of_series* and *character* were assigned weight 0.1. The remaining properties (e.g. *has_actor*, *has_director*, etc.) were assigned weight 0.8. Note that, despite considering the semantics of the properties, Weighted PageRank still has a negligible average precision for TV series since the number of episode links are much higher than the number of *references-like* properties.

The traditional Eigenvector centrality did not converge in 200 iterations.

Finally, consider *InfoRank I*, *InfoRank II*, and *InfoRank III* with the number of top neighbors set to $z = 10$, $z = 100$ and $z = 1000$ (we refer to these variations as *InfoRank III - z*). We stress that *InfoRank I* simply counts the number of datatype properties and can be computed on-the-fly by a SPARQL query. Observe that:

- *InfoRank III - 1000* has the best average precision for movies
- *InfoRank III - 100* has the best average precision for TV series
- *InfoRank I* has the best average precision for actors
- *InfoRank I* has the best combined MAP and can be computed efficiently

We now turn to a different form of performance evaluation. We argue that the IMDb movies list better capture the idea of notorious films rather than popular films, and the latter is what we are looking for. To give an example, some highly voted movies, such as Star Wars Episode V (978,712 votes), Star Wars Episode VI (804,692 votes) and Spider-Man (604,123 votes), do not appear in the IMDb movies list. So, they are not considered as a relevant movie in the evaluation of Table 2. Hence, to capture the notion of popularity, we propose to also evaluate the measures using a simple sum of the number of users' votes, instead of the movie rates.

Furthermore, note that, in these experiments, we compared only the instances of classes Movies, Series and Actors with the gold standard lists from IMDb. However, when we consider all instances from all classes, we argue that, in an IMDb dataset, popular movies should have higher scores than instances from other classes, like actors, characters, producers, etc. Therefore, in Table 3, we present the Top 5 instances obtained by the centrality and node importance measures, with the respective number of votes from IMDb users.

We note that, in traditional PageRank, the top 5 instances are highly connected nodes, such as the character names Himself/Herself, which is the name given to a character when the actor/actress plays himself/herself in a work. A similar scenario happens with *In-Degree*, *Degree* and *Weighted PageRank* measures. The *Out-degree* has a more reasonable result, however it is still influenced by all types of properties, such as the movie named "Lemonade: Detroit", which has 1,730 producers.

In the case of *InfoRank I*, the top 5 instances are from classes Series and Actress, which is a better result than *PageRank*, but they are not yet what we expected (e.g. Dolly Parton is actually better known as a singer). Finally, *InfoRank III - Top 1000* presents only highly voted movies, which it was what we expected as a result of a ranking of instances from an IMDb RDF Graph.

**Table 3**. Overall Top 5 Instances.

| PageRank | | |
|---|---|---|
| **Instance** | **Class** | **Votes** |
| Himself | Character | - |
| Herself | Character | - |
| Himself - Host | Character | - |
| Herself - Host | Character | - |
| Herself - Hostess | Character | - |
| | Total | - |
| **Out-degree** | | |
| **Instance** | **Class** | **Votes** |
| Star Wars | Movie | 1,050,455 |
| Wizard of Oz | Movie | 331,894 |
| Lemonade: Detroit | Movie | 32 |
| Star Trek | TV Series | 61,389 |
| Around the World in 80 Days | Movie | 20,662 |
| | Total | 1,464,432 |
| **InfoRank I** | | |
| **Instance** | **Class** | **Votes** |
| The Amazing Race | TV Series | 12,462 |
| Dolly Parton | Actress | - |
| The Lord of the Rings I | Movie | 1,408,061 |
| Titanic | Movie | 891,580 |
| The Dark Knight | Movie | 1,918,159 |
| | Total | 4,226,818 |
| **InfoRank III - Top 1000** | | |
| **Instance** | **Class** | **Votes** |
| Star Wars | Movie | 1,050,455 |
| The Lord of the Rings I | Movie | 1,407,013 |
| Wizard of Oz | Movie | 331,894 |
| Star Wars Episode V | Movie | 978,712 |
| Titanic | Movie | 891,580 |
| | Total | 4,659,301 |

## 6    Conclusions and Future Work

In this work, we proposed novel node importance measures for degree-decoupled RDF graphs, since the majority of existent measures (mostly adapted from PageRank) highly depend on the node degree. Our strategies, named *InfoRank I, InfoRank II and InfoRank III*, are combinations of three intuitions: (I) important things have lots of information about them; (II) important things are surrounded by other important things; (III) few good friends are better than many acquaintances.

For Intuition I, we observed that, as a consequence of the way that RDF graphs are typically modeled, important nodes usually have more datatype properties (information) than less important nodes. Intuition II is the essence of PageRank. However, we tried to improve it with Intuition III. Since we want to focus on the quality of the neighbors rather than on their quantity, we favored an approach that captures the notion that "few popular neighbors (or friends) are better than many unpopular neighbors".

*InfoRank* I uses only *Intuition I*, that is, a simple count of the datatype properties of the nodes. *InfoRank II* is an adaptation of Eigenvector Centrality using the count of datatype properties as node weights; hence, InfoRank II uses *Intuitions I* and *II*. Finally, *InfoRank III* further improves *InfoRank II* with *Intuition III* by propagating the importance only of the top neighbors of a node when using Eigenvector Centrality.

We evaluated our node importance measures in an IMDb dataset transformed to RDF. Our goal was to analyze if, using a measure based only in the structure of the graph, we could achieve a good popularity score for instances. In the first evaluation, we compared the ranks of some measures (e.g. Degree, PageRank, InfoRank, etc.) with gold standard lists for movies, TV series and actors based on user ratings, extracted from the IMDb Web site. Furthermore, since our goal was to capture a notion of popular works rather than notorious ones, we performed a second evaluation using the number of users' votes as a measure of popularity. In the first evaluation, we showed that *InfoRank I* outperforms other measures, on average. We highlight that using only *InfoRank I* is good indicator of importance and is very efficient, since it does not require the Power Iteration method, as PageRank or *InfoRank II* and *III*, and can be computed on-the-fly using a simple SPARQL query. In the second evaluation, we showed that *InfoRank III - Top 1000*, that is, using only the top 1000 neighbors to propagate the importance, outperforms other measures considering movie popularity.

As future work, we intend to use the proposed node importance measures to improve a keyword search engine over RDF graphs and to test them in other domains. It would also be interesting to test variations of InfoRank III using other top neighbor parameters.